\documentclass[%
 amsmath,amssymb,
 aip,pop
]{revtex4-2}

\usepackage{graphicx}
\usepackage{bm}

\usepackage{xcolor}

\begin{document}

\preprint{}

\title{Evolution of Field Line Helicity in Magnetic Relaxation}

\author{A. R. Yeates}
 \email{anthony.yeates@durham.ac.uk}
\affiliation{
Department of Mathematical Sciences, Durham University, Durham, DH1 3LE, UK
}
\author{A. J. B. Russell}%
 \email{a.u.russell@dundee.ac.uk}
\affiliation{
Mathematics, School of Science \& Engineering, University of Dundee, Dundee, DD1 4HN, UK
}
\author{G. Hornig}%
 \email{g.hornig@dundee.ac.uk}
\affiliation{
Mathematics, School of Science \& Engineering, University of Dundee, Dundee, DD1 4HN, UK
}

\date{\today}

\begin{abstract}
Plasma relaxation in the presence of an initially braided magnetic field can lead to self-organization into relaxed states that retain non-trivial magnetic structure. These relaxed states may be in conflict with the linear force-free fields predicted by the classical Taylor theory, and remain to be fully understood.
Here, we study how the individual field line helicities evolve during such a relaxation, and show that they provide new insights into the relaxation process.
The line helicities are computed for numerical resistive-magnetohydrodynamic simulations of a relaxing braided magnetic field with line-tied boundary conditions, where the relaxed state is known to be non-Taylor.
Firstly, our computations confirm recent analytical predictions that line helicity will be predominantly redistributed within the domain, rather than annihilated.
Secondly, we show that self-organization into a relaxed state with two discrete flux tubes may be predicted from the initial line helicity distribution.
Thirdly, for this set of line-tied simulations we observe that the sub-structure within each of the final tubes is a state of uniform line helicity. This uniformization of line helicity is consistent with Taylor theory applied to each tube individually. However, it is striking that the line helicity becomes significantly more uniform than the force-free parameter.
\end{abstract}

\keywords{}

\maketitle

\section{Introduction}

Magnetic fields in plasmas spontaneously self-organize into lower-energy states, for example powering disruptions in laboratory plasma and fusion devices \citep{Taylor1986,Taylor2000}, or stellar flares \citep{HeyvaertsPriest1984,BrowningPriest1986}. Understanding the resulting states produced by these dynamical relaxation events is crucial to understanding energy release in these systems. This is a challenging problem because the relaxation process is typically turbulent.

The most influential theory of this process, known as Taylor relaxation, proposes that the plasma minimizes energy while conserving its global magnetic helicity \citep{Taylor1974,Montgomery1978}. 
Helicity is an important invariant in magnetohydrodynamics (MHD); physically, it quantifies the average linking between magnetic field line curves in 3-dimensional (3D) space \citep{Moffatt1969}. Classic Taylor relaxation treats the global magnetic helicity as the only topological constraint on self-organization of the plasma, and hence predicts that the resulting magnetic field ${\bf B}$ is a linear force-free field obeying $\nabla\times{\bf B}=\lambda_0{\bf B}$ where $\lambda_0$ is a constant \citep{Taylor1974,Taylor1986,Taylor2000,Woltjer1958}. The underlying assumption that the total magnetic helicity is more highly conserved than energy, is an example of selective decay in turbulent fluids and plasmas \citep{MatthaeusMontgomery1980}. 

Taylor's theory successfully reproduces major features of the magnetic field in a reversed field pinch, which it was developed to explain, and it has given important insights into other experimental plasma physics phenomena such as current limitation and symmetry breaking \citep{Taylor2000}. It has also been applied to astrophysical plasmas such as magnetized jets \citep{KoeniglChoudhuri1985}, interplanetary magnetic clouds \citep{Burlaga1988,KumarRust1996} and solar flares  \citep{HeyvaertsPriest1984,BrowningPriest1986,Nandy2003}. However, the theory is less successful in unbounded configurations and tokamaks\citep{Bhattacharjee1982,YeeBallan2000}. 
Indeed, since the Taylor state is force-free, with ${\nabla p = 0}$ in the absence of gravity, it cannot (in its classic form) explain any self-organized field that magnetically confines plasma.
Recently, 3D MHD simulations with ever-increasing Lundquist numbers have made it possible to probe turbulent relaxation in detail, at low cost, and for precisely known initial conditions and parameters. These numerical experiments have found more counterexamples where the end states are not linear force-free fields \citep{Amari2000,WilmotSmith2009,Yeates2010,Yeates2015,Bareford2013}. It therefore appears more certain than ever before that other constraints, in addition to the global helicity, can be important for plasma self organization.

Previous efforts to generalize Taylor relaxation have considered additional global constraints on the magnetic topology, such as the topological degree of the field line mapping \citep{Yeates2010,Yeates2015}, measures of higher-order linking \citep{Evans1992,HornigMayer2002,Akhmetev2004}, and helicity integrals weighted by powers of the helical flux function \citep{Bhattacharjee1980}.
Alternative relaxation models have also been developed where the constraints are not purely due to the initial magnetic topology\citep{1981Turner, Hudson2012}, as well as models where flows occur in the relaxed state \citep{Bates1998}.
In this paper, we take a novel approach by examining the distribution of individual field line helicities \citep{Berger1988,YeatesHornig2011} in a turbulent MHD relaxation in a straight geometry. In his original paper, Taylor \citep{Taylor1974} noted that a helicity integrand exists for every magnetic field line. He also conjectured that during reconnection ``the effect of the topological changes is merely to redistribute the integrand among the field lines involved,'' and so proposed that the ``final state of relaxation, therefore, will now be the state of minimum energy subject only to the single [global] invariant.''

In the forty-seven years since Taylor's original paper, understanding of field line helicities has advanced significantly \citep{Berger1988,YeatesHornig2011,YeatesHornig2013,PriorYeates2014}, and it has recently been found that they do not evolve arbitrarily during reconnection but instead obey an evolution equation derived by Russell \textit{et al.} \citep{Russell2015}. These authors considered how the field line helicities would evolve during localized reconnection in a complex, 3D magnetic field, such as would arise during turbulent evolution of a highly conducting plasma. They derived an evolution equation for the field line helicity (to be discussed in Section~\ref{sec:rea} below). This includes both a resistive dissipation term and a ``work-like'' term, both of which involve field line integrated quantities. The authors showed that a sufficiently complex field line mapping produces a scale separation between the two terms, with the work-like term expected to dominate. 
Furthermore, the work-like term is expected to conserve the overall field line helicity to leading order, acting primarily to redistribute rather than destroy it. 
At the time, these analytic predictions were validated using kinematic examples.
In principle, such laws governing the time evolution of line helicities could lead to a conflict with the basic Taylor assumption that global helicity is the only dynamically relevant invariant. Our aim in this paper is therefore to explore how the field line helicity evolves in full 3D MHD simulations of magnetic relaxation.

 Our first objective is to test the prediction of Russell \textit{et al.} \citep{Russell2015} that field line helicity is redistributed rather than destroyed during turbulent relaxation. However, because we select a configuration known to relax to a force-free field that is not the linear force-free Taylor state \citep{Pontin2011}, we are also able to investigate whether the field line helicities shed further light on what determines the relaxed state. The paper is organized as follows. The numerical simulations are described in Section~\ref{sec:sims}, which includes a summary of our three main observations about magnetic relaxation and field line helicity (Section~\ref{sec:obs}). Each of these three observations is then discussed in more detail in a separate section: Section~\ref{sec:rea} investigates the extent to which field line helicity is redistributed rather than destroyed; Section~\ref{sec:top} examines the overall topology of the final state and its relation to the topology of the initial line helicity distribution; and Section~\ref{sec:uni} considers the finer sub-structure of the relaxed state. 
 Conclusions are given in Section~\ref{sec:con}. 

\section{Numerical simulations} \label{sec:sims}

MHD turbulent relaxation can be simulated by solving the resistive MHD equations with low dissipation and a braided initial magnetic field. The choice of a braided field is motivated by the significant turbulence that these fields generate, an important prerequisite for Taylor relaxation.

\subsection{Resistive-MHD equations and parameters}

We used the Lare3d Lagrangian-remap code \cite{Arber2001} to solve the resistive-magnetohydrodynamic (MHD) equations in a Cartesian domain $[-8,8]\times[-8,8]\times[-24,24]$. The code solves the non-dimensionalized equations
\begin{align}
    \frac{\partial\rho}{\partial t} &= -\nabla\cdot(\rho{\bf v}),\\
    \rho\frac{\mathrm{D}{\bf v}}{\mathrm{D}t} &= {\bf j}\times{\bf B} - \nabla p + \nabla\cdot\bm{\sigma},\\
    \frac{\partial{\bf B}}{\partial t} &= \nabla\times({\bf v}\times{\bf B}) - \nabla\times(\eta{\bf j}), \label{eqn:induct}\\
    \rho\frac{\mathrm{D}\epsilon}{\mathrm{D}t} &= -p\nabla\cdot{\bf v} + \eta|{\bf j}|^2 + \bm{\epsilon}:\bm{\sigma},\\
    p &= \rho\epsilon(\gamma - 1),\\
    \mu_0{\bf j} &= \nabla\times{\bf B},
\end{align}
where $\rho$ is the mass density, ${\bf v}$ the plasma velocity, ${\bf B}$ the magnetic field, ${\bf j}$ the current density, $p$ the plasma pressure, $\bm{\sigma}$ the stress tensor, $\epsilon$ the specific internal energy density, $\eta$ the resistivity, $\bm{\epsilon}$ the strain tensor and $\gamma=\tfrac53$ the ratio of specific heats. The viscous term $\nabla\cdot\bm{\sigma}$ includes only a shock viscosity to prevent unphysical oscillations, but no background viscosity. The shock viscosity takes the tensor form given by \citet{Bareford2013}, with the same dimensionless parameters $\nu_1=0.1$ and $\nu_2=0.5$. There is a corresponding heating term $\bm{\epsilon}:\bm{\sigma}$ in the energy equation. The resistivity $\eta$ is uniform with no enhancement at current sheets. The non-dimensional time $t$ is such that one unit is the time taken for an Alfv\'en wave when $|\mathbf{B}|=\rho=1$ to travel a unit distance.

Four different runs are illustrated in this paper, differing in Lundquist number, $S$, as listed in Table \ref{tab:runs}. Since the non-dimensional $t$ represents the Alfv\'en time, the Lundquist number is simply given by $S=\eta^{-1}$, in terms of the non-dimensional resistivity $\eta$.

\begin{table}
\caption{\label{tab:runs} List of simulation runs.}
\begin{ruledtabular}
\begin{tabular}{lll}
Lundquist number [$S$] & Diffusivity [$\eta$] & Grid resolution  \\
\hline
         2500 & $4\times10^{-4}$ & $320\times 320\times 240$\\
         5000 & $2\times10^{-4}$ & $640\times 640\times 480$\\
         10000 & $1\times10^{-4}$ & $640\times 640\times 480$\\
         20000 & $5\times10^{-5}$ & $960\times 960\times 720$\\
\end{tabular}
\end{ruledtabular}
\end{table}

\subsection{Boundary and initial conditions}

For the simulations described here, line-tied boundary conditions were implemented by setting the plasma velocity ${\bf v}={\bf 0}$ on all six boundaries, with zero normal-gradient conditions for ${\bf B}$, $\rho$, and $\epsilon$. Such conditions are relevant to coronal loops and laboratory experiments such as the large plasma research device \citep{Gekelman1991}. Periodic boundary conditions -- as relevant to laboratory fusion devices -- can also be considered, but we defer their discussion to a subsequent paper.

The simulations were initialized with uniform $\rho=1$ and $\epsilon=0.01$. The initial magnetic field took the ``braided'' E3 form \cite{WilmotSmith2009}, which consists of six twists of magnetic flux superimposed on a uniform magnetic field, arranged in two offset columns. This field has a global helicity of zero and its dynamics have been extensively investigated in a long-term research programme \citep{Pontin2016}. Explicitly,
\begin{align}
    B_x(x,y,z) &= -\sqrt{2}\sum_{i=1}^6 k_i y \exp\left(-\frac{\xi_i^2}{4} \right),\label{eqn:b01}\\
    B_y(x,y,z) &= \sqrt{2}\sum_{i=1}^6 k_i (x - x_i) \exp\left(-\frac{\xi_i^2}{4} \right),\label{eqn:b02}\\
    B_z(x,y,z) &= 1, \label{eqn:b03}
\end{align}
where $\xi^2_i = 2(x-x_i)^2 + 2y^2 + (z-z_i)^2$ and $x_i=k_i=(1,-1,1,-1,1,-1)$ and $z_i=(-20,-12,-4,4,12,20)$.

The plasma beta of this initial condition is $\beta\approx 0.01$. There are initially significant unbalanced ${\bf j}\times{\bf B}$ forces, so the dynamical evolution begins immediately. However, previous simulations \cite{Yeates2010} found consistent relaxed states whether or not the magnetic field was first relaxed using an ideal Lagrangian code.

\subsection{Overview of the evolution}

\begin{figure*}
    \centering
    \includegraphics[width=\textwidth]{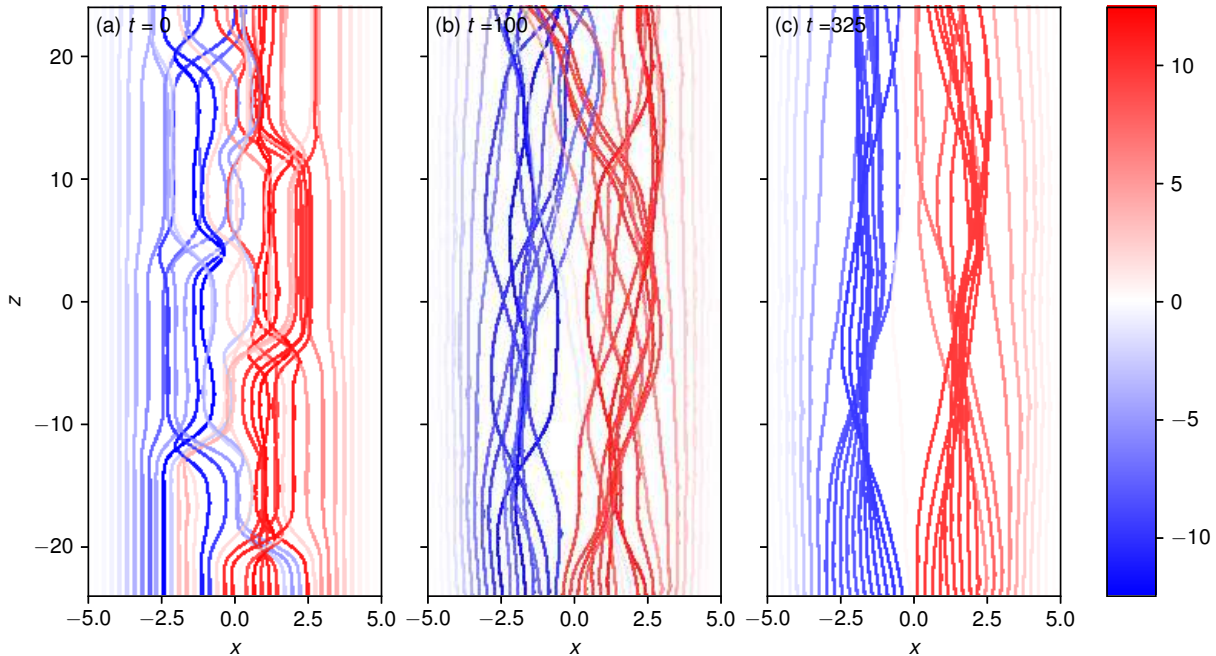}
    \caption{Magnetic field lines for $S=10\,000$, (a) in the initial braided field, (b) during the turbulent relaxation, and (c) after the dynamic phase. The field lines are traced from fixed locations with $y=0$ on the $z=-24$ boundary, and their intersections with $z=24$ change substantially as the magnetic field simplifies into a pair of oppositely-twisted flux tubes. Field lines are colored by field line helicity, $\mathcal{A}$, and shown in planar projection.
    }
    \label{fig:3D_vis}
\end{figure*}

Figure~\ref{fig:3D_vis} provides an overview of the simulation, starting from the initial condition (Fig.~\ref{fig:3D_vis}a). In a brief starting phase, the initial magnetic twists launch Alfv\'en waves that interact non-linearly to generate MHD turbulence. Thereafter (Fig.~\ref{fig:3D_vis}b), the plasma and magnetic field evolve rapidly as magnetic field lines reconnect in the turbulent braided region. The outcome (Fig.~\ref{fig:3D_vis}c) is a simpler lower-energy magnetic field consisting of a pair of oppositely twisted flux tubes. These two flux tubes co-exist stably, and running the simulation for longer produces only a gradual resistive diffusion of the magnetic field with no further simplification of the overall topology.

\begin{figure}
    \centering
    \includegraphics[width=0.8\textwidth]{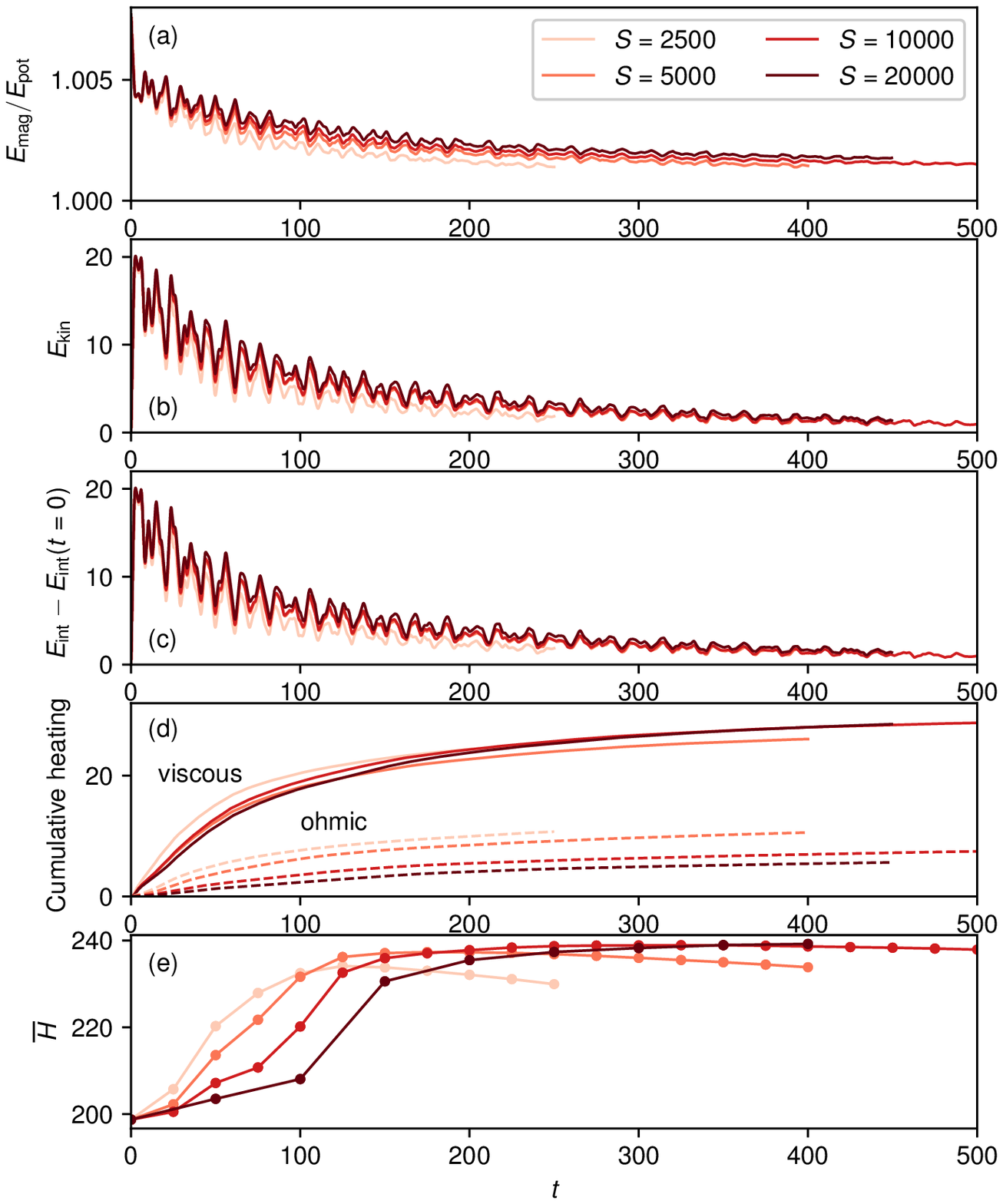}
    \caption{Time evolution of integrated quantities: (a) magnetic energy relative to the potential field ${\bf B}={\bf e}_z$; (b) kinetic energy; (c) excess internal energy; (d) cumulative heating rates, and (e) unsigned helicity, $\overline{H}$.}
    \label{fig:diagnostics}
\end{figure}

\begin{figure}
    \centering
    \includegraphics[width=0.6\textwidth]{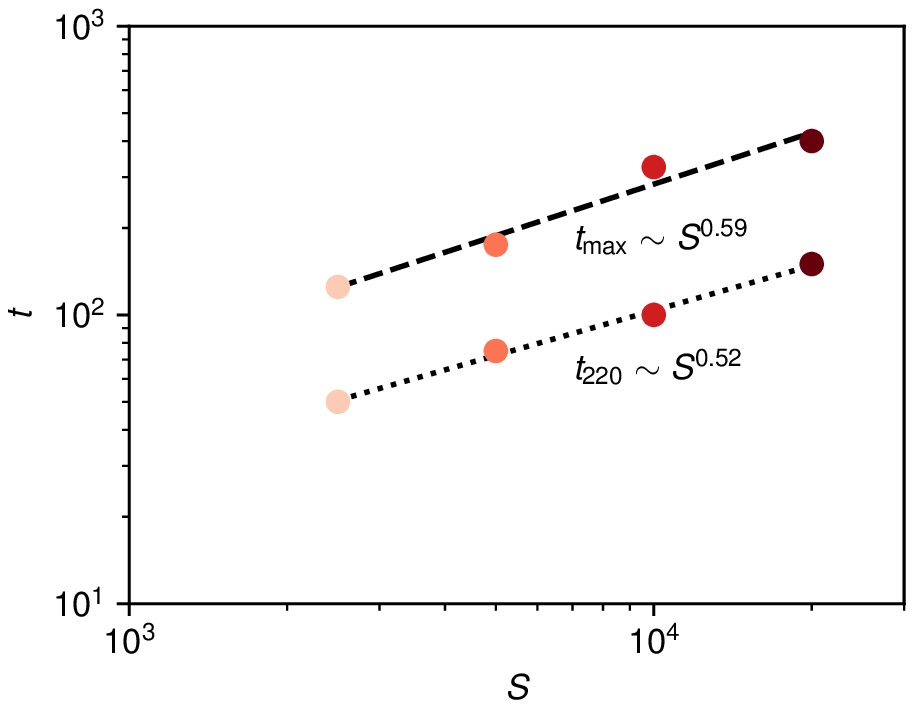}
    \caption{Dependence of the relaxation time on Lundquist number, $S$, as measured by the evolution of $\overline{H}$ (equation \eqref{eqn:hbar}). The two points $t_{220}$ and $t_{\rm max}$ for each simulation run indicate respectively the first snapshot where $\overline{H}>220$ -- representing approximately half of its overall change -- and the snapshot with maximum $\overline{H}$. The dotted and dashed lines are least squares fits giving the indicated scalings.}
    \label{fig:timescale}
\end{figure}

The time evolution of various integrated quantities is shown in Figure~\ref{fig:diagnostics}. All simulations show a clear relaxation in terms of magnetic, kinetic and internal energies (Figs. \ref{fig:diagnostics}a-c, respectively). The oscillations visible in the energies are ideal in origin and their frequency is independent of $S$. Their period (of the order 10 time units) is consistent with the Alfv\'en travel time between the initial flux rings.
However, the non-ideal evolution which enables the relaxation to take place has an overall timescale which takes longer for larger $S$. Figure~\ref{fig:timescale} shows that the relaxation time scales approximately as $S^{0.5}$ to $S^{0.6}$. For this reason, we ran the simulations with larger $S$ for longer. Notice in Figure~\ref{fig:diagnostics}(d) that the net viscous heating increases with $S$, commensurate with the higher kinetic energy, whereas the net ohmic heating decreases with $S$.

\subsection{Observed behavior of field line helicity} \label{sec:obs}

We have analyzed the evolution of field line helicity in the numerical simulations. Without magnetic reconnection, not only would the total magnetic helicity be conserved, but so would the field line helicity,
\begin{equation}
    \mathcal{A}(L) = \int_L{\bf A}\cdot\,{\rm d}{\bf l},
\end{equation}
of every individual magnetic field line $L$. Here ${\bf A}$ is a vector potential such that ${\bf B}=\nabla\times{\bf A}$. Physically, $\mathcal{A}(L)$ measures the winding of magnetic flux around the field line of interest \citep{YeatesHornig2011,PriorYeates2014}, so its ideal invariance follows from field line conservation (Alfv\'en's theorem), together with fixing ${\bf n}\times{\bf A}$ at the boundaries. This fixing is possible because the line-tying leaves $B_n$ stationary on all six boundaries \citep{Russell2015, Yeates2018}. We set ${\bf n}\times{\bf A} = {\bf n}\times{\bf A}^{\rm ref}$  where ${\bf A}^{\rm ref}=\left(-\tfrac12y, \tfrac12x, 0\right)$ is a vector potential for the current-free reference field ${\bf B}^{\rm ref}={\bf e}_z$ that satisfies $B^{\rm ref}_n=B_n$ on all six boundaries. Strictly speaking, our initial condition \eqref{eqn:b01}-\eqref{eqn:b03} does not precisely satisfy this condition on the side boundaries, but the domain is sufficiently large that the maximum error in $B_n$ is of the order $10^{-10}$, much smaller than the truncation errors in our numerical calculations. Our method for computing ${\bf A}$ is described in Appendix \ref{app:a}.

In our resistive relaxation, the distribution of $\mathcal{A}$ can change due to reconnection \cite{Russell2015}. In fact, Yeates \& Hornig \cite{YeatesHornig2013} proved it will necessarily do so if there is any change in the field line connectivity between the two end boundaries. This is the redistribution of helicity invoked by Taylor \citep{Taylor1974}. Since $\mathcal{A}$ provides a complete description of the magnetic field connectivity, it is a natural tool for studying the evolution of the magnetic field structure.

To compute $\mathcal{A}$, we first computed appropriate vector potentials, ${\bf A}$, for a sequence of simulated snapshots of ${\bf B}$, using the method described in Appendix \ref{app:a}. For each snapshot, we traced magnetic field lines from a regular grid of $1024\times 1024$ starting points in the region $\{-4\leq x\leq 4, -4\leq y\leq 4\}$ on the lower boundary $z=-24$, using a second-order Runge-Kutta method with adaptive step-size. We focus on this smaller region where the reconnection takes place because $\mathcal{A}$ remains unchanged outside. We then integrated ${\bf A}\cdot{\bf B}/|{\bf B}|$ along each field line with the composite trapezium rule. Trilinear interpolation was used for both ${\bf A}$ and ${\bf B}$.

\begin{figure}
    \centering
    \includegraphics[width=0.9\textwidth]{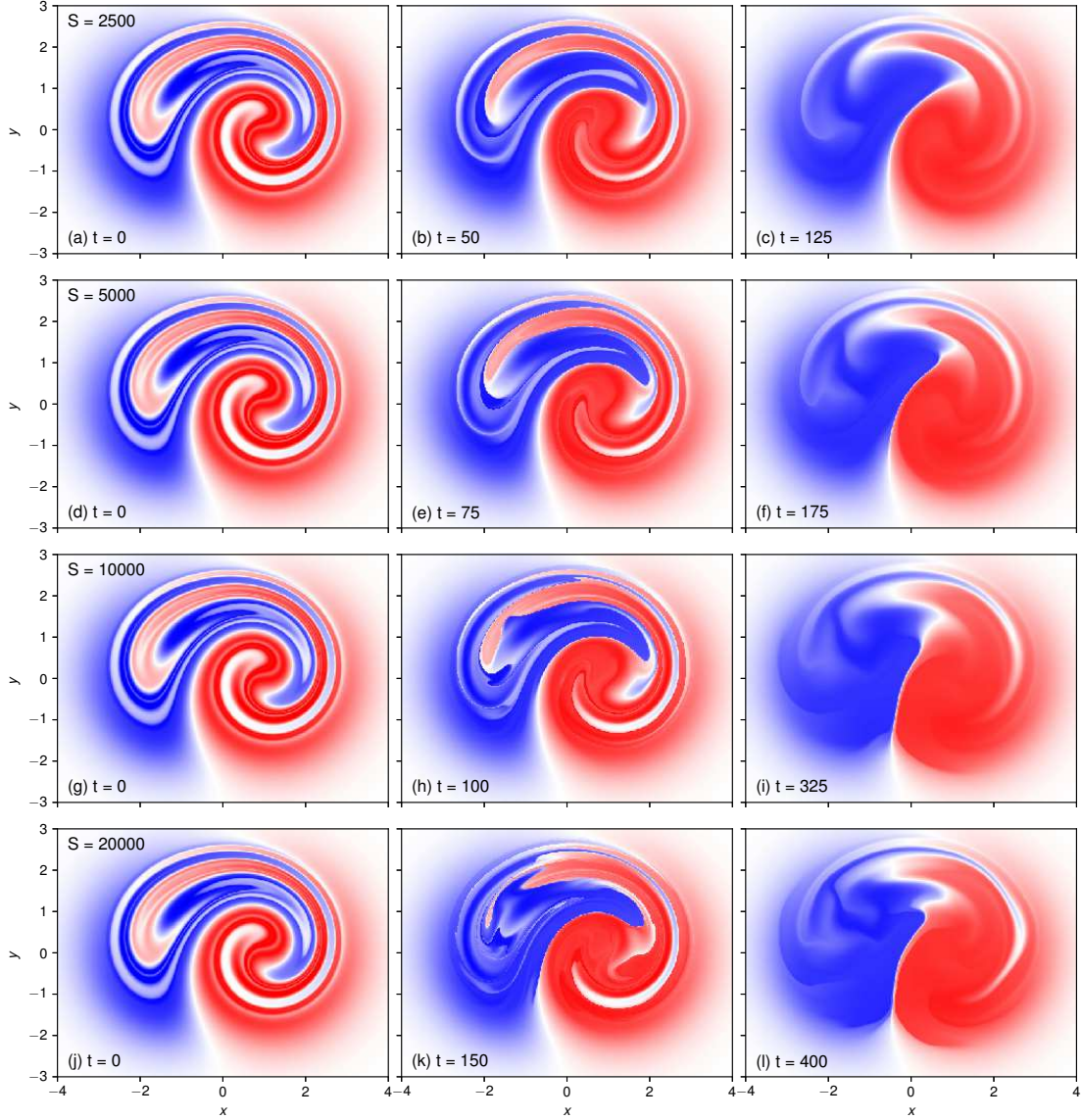}
    \caption{Colormaps of $\mathcal{A}$ for the simulation runs with different $S$, in the initial configuration (a, d, g, j), at time $t=t_{220}$ during the turbulent dynamics (b, e, h, k) and at the end of this phase at time $t=t_{\rm max}$ (c, f, i, l). The field-line helicity is plotted on the $z=-24$ cross-section, with the red/blue color scale capped at $\mathcal{A}=\pm 12$ in all cases. The times $t_{220}$ and $t_{\rm max}$ are chosen according to $\overline{H}$ as in Fig.~\ref{fig:timescale}.}
    \label{fig:flh2d}
\end{figure}

Figure~\ref{fig:flh2d} shows the evolution of $\mathcal{A}$, which, we stress, changes only by magnetic reconnection. The times in this figure are chosen to be approximately equivalent between runs, in terms of the progress of the topological evolution. This is most conveniently measured by the unsigned helicity,
\begin{equation}
        \overline{H}=\int_{-4}^4\int_{-4}^4|\mathcal{A}B_z|\,\mathrm{d}x\mathrm{d}y,
                \label{eqn:hbar}
\end{equation}
which follows a similar path in all simulations, as shown in Fig.~\ref{fig:diagnostics}(e).
We make three main observations:
\begin{enumerate}
    \item There is no wholesale annihilation of line helicity, despite the fact that annihilation would be consistent with conservation of the global helicity, which is equal to zero. In fact, the mean value of $|\mathcal{A}|$ in the region $\{-4\leq x\leq 4, -4\leq y\leq 4\}$ increases by approximately 20\%. This is shown by the evolution of $\overline{H}$ in Fig.~\ref{fig:diagnostics}(e). The absence of annihilation accords with the expectation from \citet{Russell2015} that the leading order evolution of line helicity in a highly-conducting fluid will be redistribution rather than dissipation. This will be discussed further in Section~\ref{sec:rea}.
    \item In the final state, the positive and negative line helicity are organized into two distinct regions, with the exception of some surviving remnants of the mixed structure on the outside of the relaxation region (toward the top in Figure~\ref{fig:flh2d}). This topological organization of the relaxed state will be discussed in Section~\ref{sec:top}.
    \item Within each of the final positive and negative regions, the line helicity is very uniform. This was not anticipated by \citet{Russell2015}, and will be discussed in Section~\ref{sec:uni}.
\end{enumerate}

\section{Dominance of redistribution} \label{sec:rea}

\begin{figure}
    \centering
    \includegraphics[width=0.8\textwidth]{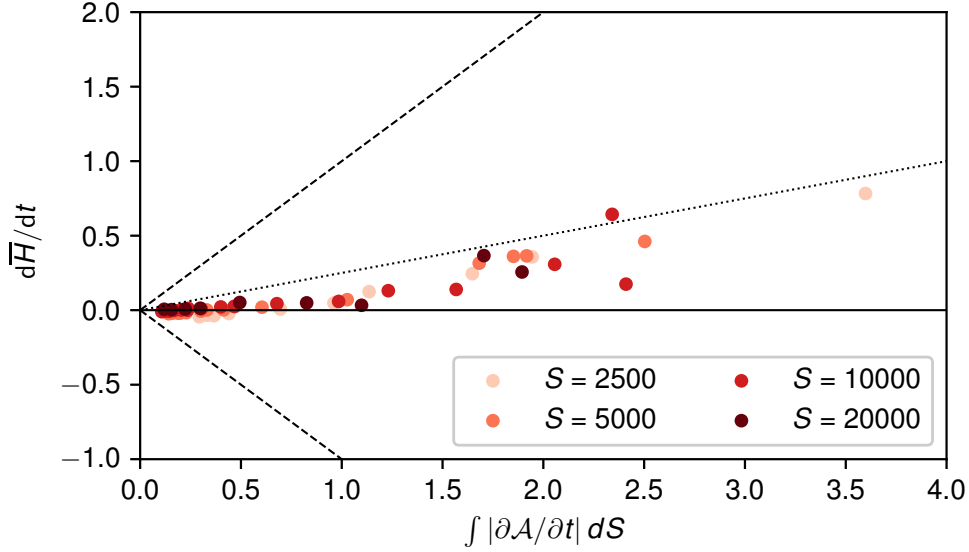}
    \caption{Scatterplot comparing the rate of change of unsigned helicity $\overline{H}$ to the integrated unsigned changes of $\mathcal{A}$. Each dot corresponds to a time snapshot as shown in Fig.~\ref{fig:diagnostics}(e). The time derivatives were computed using two $1024\times 1024$ maps of $\mathcal{A}$ taken at short intervals $\Delta t\sim 0.1$, and the integrals were taken over the region $-4\leq x\leq 4$, $-4\leq y\leq 4$. Dashed lines indicate slope $\pm1$ and the dotted line indicates slope $0.25$.}
    \label{fig:derivs}
\end{figure}

We have seen that wholesale annihilation of $\mathcal{A}$ does not take place during the simulation, as evidenced by the fact that the overall unsigned helicity $\overline{H}$ does not decrease (Figure~\ref{fig:diagnostics}e). But we have also seen, in Figure \ref{fig:flh2d}, that the $\mathcal{A}$ pattern evolves significantly during the relaxation, so there must nevertheless be significant local changes in $\mathcal{A}$. To  quantify the level of these local changes, Figure~\ref{fig:derivs} compares the net rate of change of $|\mathcal{A}|$, as measured by $\mathrm{d}\overline{H}/\mathrm{d}t$, with the integrated local rates of change $|\partial\mathcal{A}/\partial t|$. The latter are at least four times higher than the rate of change of $\overline{H}$, so that the majority of local changes in $\mathcal{A}$ are cancelled out globally.
This accords with the dominance of pairwise increases and decreases predicted by  \citet{Russell2015}. These authors proposed that the dominance of redistribution of $\mathcal{A}$ over its dissipation in a highly-conducting plasma may be explained by the evolution equation for $\mathcal{A}$. Next, we compute the terms in this equation to illustrate that this basic explanation holds in our simulations. 

\subsection{Line helicity evolution equation}

\citet{Russell2015} showed that the line helicity of a field line $L$ evolves according to 
\begin{equation}
   \frac{\mathrm{d}\mathcal{A}(L)}{\mathrm{d} t} = \Big[{\bf w}\cdot{\bf A} - \Psi\Big]_{{\bf x}_0}^{{\bf x}_1},
   \label{eqn:dafull}
\end{equation}
where ${\bf x}_0$ and ${\bf x}_1$ are the endpoints of $L$, and ${\bf w}$ is any field line velocity into which $L$ is frozen. The last term represents the voltage drop along the field line,
\begin{equation}
    \Big[\Psi\Big]_{{\bf x}_0}^{{\bf x}_1} = \int_L\eta{\bf j}\cdot\,\mathrm{d}{\bf l}.
\end{equation}
Since $B_z>0$ everywhere, $\Psi$ is global and single-valued. There is no additional scalar potential (i.e. gauge) term in Equation~\eqref{eqn:dafull} because its contributions mutually cancel due to a combination of (i) our gauge restriction fixing ${\bf n}\times{\bf A}$ on the end boundaries, and (ii) the line-tied boundary condition that both ${\bf v}$ and ${\bf n}\times{\bf j}$ vanish on these boundaries.

To compare with the simulations shown in Figure~\ref{fig:flh2d}, it is beneficial to choose ${\bf w}$ in Equation~\eqref{eqn:dafull} so that the left-hand side represents the change in $\mathcal{A}$ at a fixed position on the $z=-24$ boundary. In other words, so that field lines are identified over time by fixed endpoints ${\bf x}_0$ on this boundary. As discussed by \citet{Russell2015}, this corresponds to
\begin{equation}
    {\bf w} = \frac{{\bf e}_z\times\big(\nabla\Psi + {\bf v}\times{\bf B} - \eta{\bf j}\big)}{B_z},\label{eqn:w}
\end{equation}
with the choice that $\Psi=0$ on the $z=-24$ boundary. It follows that ${\bf w}={\bf 0}$ on this boundary, so that field lines are traced from fixed endpoints there. But on the far boundary at $z=24$, we then have ${\bf e}_z\times\nabla\Psi \neq {\bf 0}$ so the opposite end-points move due to reconnection. With this ${\bf w}$, Equation~\eqref{eqn:dafull} reduces to
\begin{equation}
       \frac{\partial\mathcal{A}({\bf x}_0)}{\partial t} = ({\bf w}\cdot{\bf A})({\bf x}_1) - \Psi({\bf x}_1),
   \label{eqn:dasimp}
\end{equation}
where ${\bf x}_0$ is fixed but the opposite end-point ${\bf x}_1$ moves in time.

\subsection{Numerical verification}

To compute the terms on the right-hand side of Equation~\eqref{eqn:dasimp} from the simulations, we first integrate $\eta{\bf j}$ along field lines to calculate $\Psi$ on the $z=24$ boundary, using a similar field line integration method to that used for calculating $\mathcal{A}$. Then, since $B_z=1$, Equation~\eqref{eqn:w} gives
\begin{equation}
    {\bf w}({\bf x}_1) = -\frac{\partial\Psi({\bf x}_1)}{\partial y_1}{\bf e}_x + \frac{\partial\Psi({\bf x}_1)}{\partial x_1}{\bf e}_y,
    \label{eqn:wdxi}
\end{equation}
and ${\bf A}({\bf x}_1) = {\bf A}^{\rm ref}({\bf x}_1)$. The derivatives are computed numerically with central differencing using field line startpoints spaced by $10^{-3}$.

\begin{figure}
    \centering
    \includegraphics[width=\textwidth]{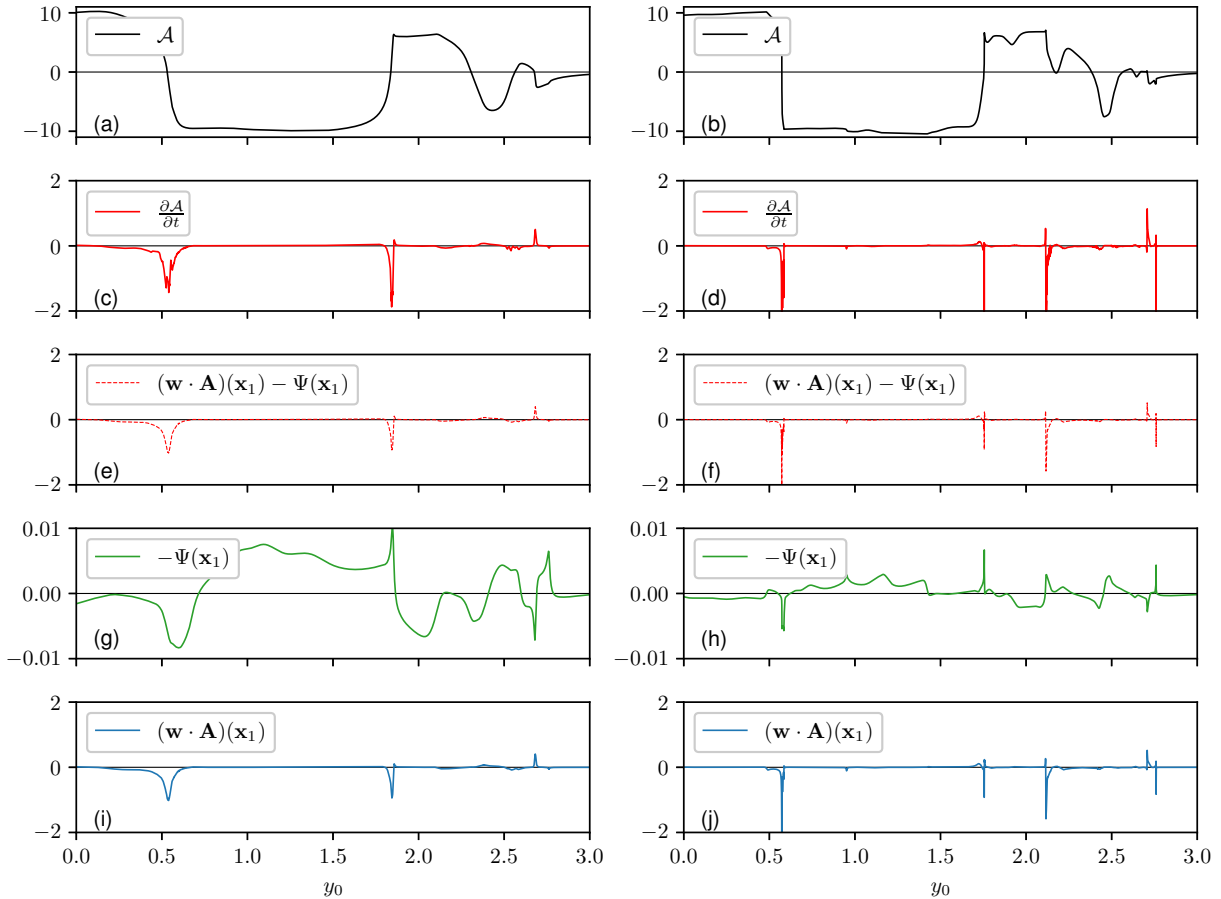}
    \caption{Terms in the evolution equations for $\mathcal{A}$, in a radial cut at $x_0=0$ on the $z=-24$ boundary. Left column shows $t=100$ for $S=2500$ run, while right column shows $t=150$ for $S=10\,000$. Panels (a/b) show $\mathcal{A}$, and (c/d) show $\partial\mathcal{A}/\partial t$ computed using forward differences with $\mathrm{d}t=0.013$ (for $S=2500$) and $\mathrm{d}t=0.006$ (for $S=10\,000$).  Panels (e/f) show the recovered right-hand side of \eqref{eqn:dasimp}, while (g-j) show the two constituent terms.}
    \label{fig:eveqn}
\end{figure}

In Figure~\ref{fig:eveqn}, we check that Equation~\eqref{eqn:dasimp} holds in  our numerical simulations. The figure shows one-dimensional cuts through two of the runs ($S=2500$ and $S=10\,000$). The times are chosen shortly before the end of the dynamical relaxation phase in each case -- between the middle and right columns of Figure~\ref{fig:flh2d}. Panels (c/d) and (e/f) in this figure show the left- and right-hand sides of Equation~\eqref{eqn:dasimp}, as computed numerically. We see good agreement despite the sharp peaks in the field-line integrated quantity $({\bf w}\cdot{\bf A})({\bf x}_1)$. Resolving the terms in Equation~\eqref{eqn:dasimp} becomes progressively more difficult for larger $S$, since increasing the Lundquist number supports steeper gradients in the field line mapping, which in turn produces steeper gradients in field line integrated quantities such as $\mathcal{A}$ and $\Psi$, and hence also in ${\bf w}$. The difference in steepness between the two runs is evident in the middle column of Figure~\ref{fig:flh2d}, and is clear in Figure~\ref{fig:eveqn}(a/b).

Two other features are visible in Figure~\ref{fig:eveqn}. Firstly, it clearly shows the uniformization of $\mathcal{A}$ within the negative region, which is visible as a flat plateau at this relatively advanced stage of the relaxation. We believe that this uniformization arises from the dynamics and cannot be predicted purely from the evolution equation \eqref{eqn:dasimp}; it will be discussed in Section~\ref{sec:uni}. Secondly, it is clear that the significant changes in $\mathcal{A}$ are arising from the ${\bf w}\cdot{\bf A}$ term, rather than from $\Psi$. Thus our simulations support the dominance of the work term over the voltage drop, as predicted by \citet{Russell2015}. The difference arises simply because ${\bf w}({\bf x}_1)$ depends on derivatives of the field line integrated quantity $\Psi$, through Equation~\eqref{eqn:wdxi}, and these derivatives are typically large owing to the short lengthscales.

It is interesting to note that the spikes in $\partial\mathcal{A}/\partial t$ correlate with locations where $\mathcal{A}$ has a high spatial gradient. These typically arise from locations where the field line mapping has large gradients, because $\mathcal{A}$ is a field-line integrated quantity. Both ${\bf w}$ and $\Psi$ are typically large at these locations, indicating that the corresponding field lines are reconnecting. 

\section{Global topology} \label{sec:top}

It is evident from Figure~\ref{fig:flh2d} that the distribution of $\mathcal{A}$ relaxes into two largely separated regions of opposite sign. In fact, we will show here that the impossibility of further simplification follows mathematically from the initial $\mathcal{A}$ pattern, together with the localization (in $x$ and $y$) of the resistive dynamics. By localization, we mean that there is an outer boundary region where  $\partial\mathcal{A}/\partial t$ remains small throughout the evolution because it is almost ideal there. This means that $\Psi\approx 0$ so the right-hand side of \eqref{eqn:dasimp} is small.

To see how the evolution of $\mathcal{A}$ is constrained, we note first that within these simulations it is a smooth function of ${\bf x}_0$ (albeit with steep gradients), that evolves continuously in time. Therefore its contours evolve continuously in time, except possibly at critical points where $\nabla\mathcal{A}={\bf 0}$. Moreover, these critical points cannot arbitrarily appear or disappear, but must do so in limited ways -- typically pairwise -- so as to preserve their net Poincar\'e index. This is +1 for an extremum (maximum or minimum) and -1 for a saddle. The only way that the overall net Poincar\'e index  can change is through movement of critical points in or out of the region. In our case, this is prevented by the invariance of $\mathcal{A}$ in the surrounding boundary region. So the net Poincar\'e index is invariant.

\begin{figure}
    \centering
    \includegraphics[width=\textwidth]{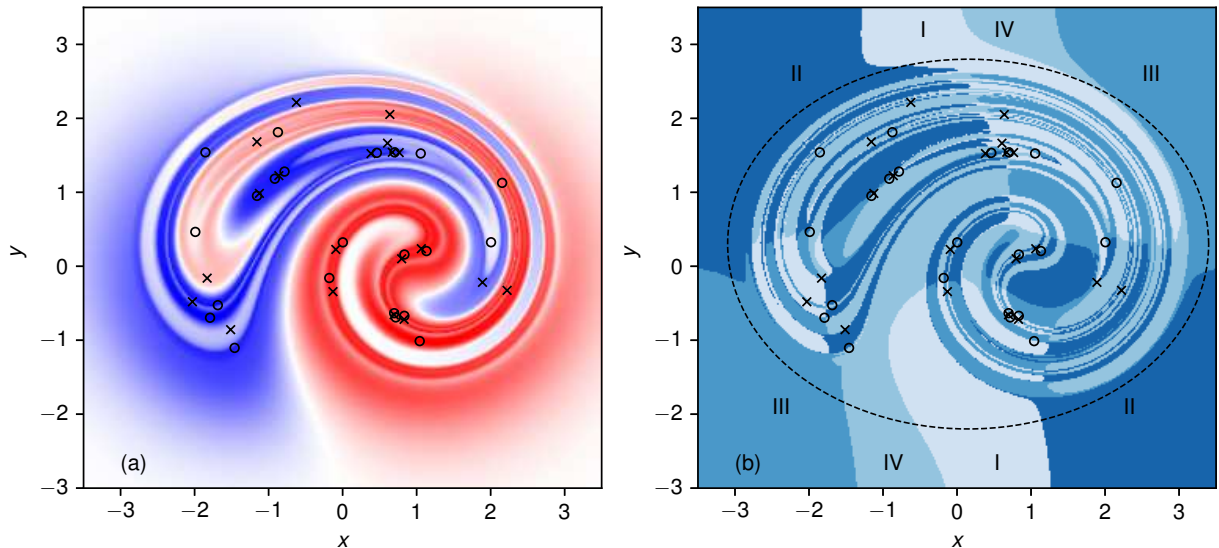}
    \caption{The topology of $\nabla\mathcal{A}$ at $t=0$. Panel (a) shows the distribution of $\mathcal{A}$ (red positive, blue negative), while panel (b) is shaded according to the orientation of $\nabla\mathcal{A}$. Specifically, (I) $\partial\mathcal{A}/\partial x>0$, $\partial\mathcal{A}/\partial y>0$; (II) $\partial\mathcal{A}/\partial x<0$, $\partial\mathcal{A}/\partial y>0$; (III) $\partial\mathcal{A}/\partial x<0$, $\partial\mathcal{A}/\partial y<0$; and (IV) $\partial\mathcal{A}/\partial x>0$, $\partial\mathcal{A}/\partial y<0$. Critical points are denoted by circles (extrema) and crosses (saddles), computed as described in Appendix \ref{app:i}. The sequence of two full counterclockwise rotations of $\nabla\mathcal{A}$ as the dashed curve is traced counterclockwise indicates that the net Poincar\'e index is two.}
    \label{fig:index}
\end{figure}

The net Poincar\'e index of the $\mathcal{A}$ pattern in our simulations has the value 2. 
Figure~\ref{fig:index} illustrates this topological structure at $t=0$. At this initial time, $\mathcal{A}$ may be computed analytically (see Appendix \ref{app:i}), allowing us to rigorously identify the 42 individual critical points in the complex pattern. These critical points are 22 extrema and 20 saddles, and the difference yields the net Poincar\'e index of 2. In fact, it is not necessary to identify individual critical points in order to compute the net Poincar\'e index: this may be determined purely from the number of rotations of the $\nabla\mathcal{A}$ vector around a single circuit of the outer ``boundary region'' (\textit{e.g.}, the dashed curve in Figure~\ref{fig:index}b). Here there are two positive rotations of $\nabla\mathcal{A}$, confirming the net value 2. It is clear from Figure~\ref{fig:flh2d} that this remains the case throughout the numerical simulations, although the total number of critical points changes, decreasing overall as the $\mathcal{A}$ pattern simplifies.

The consequence of this invariant overall Poincar\'e index of 2 is that the relaxed state must contain at least 2 extrema. One of these is a maximum and one a minimum, thus explaining the persistence of two separate regions of oppositely signed $\mathcal{A}$. This topological structure is not predicted by the standard Taylor theory, since conservation of total helicity alone would not prevent relaxation to a uniform straight magnetic field.

A topological explanation for the persistence of two tubes in these simulations was already given by \citet{Yeates2010} (see also \citet{Yeates2015}). However, that work considered the topological degree of the field line mapping from one end boundary to the other, rather than the $\mathcal{A}$ pattern. The two are related to some extent, since $\mathcal{A}$ contains all of the information about the field line mapping \citep{YeatesHornig2013}. But fixed points of the field line mapping are not, in general, critical points of $\mathcal{A}$, and \textit{vice versa}; in this case there are 22 fixed points in the initial configuration compared to 42 critical points of $\mathcal{A}$. Admittedly, the net topological degree of the mapping is 2, matching the overall Poincar\'e index of $\nabla\mathcal{A}$, but we have no reason to believe that this holds for all braided magnetic fields.

We remark that the existence of a surrounding ideal region imposes stronger constraints than merely the preservation of the net Poincar\'e index. For example, in Figure~\ref{fig:flh2d} it is clear that there is a peripheral region where a vestige of the initial mixed pattern of $\mathcal{A}$ survives, because those field lines have undergone insufficient reconnection. Nevertheless, the Poincar\'e index of the initial distribution of $\mathcal{A}$ suffices to predict the principal topological feature of two regions with opposite $\mathcal{A}$.

\section{Uniformization} \label{sec:uni}

\begin{figure}
    \centering
    \includegraphics[width=0.6\textwidth]{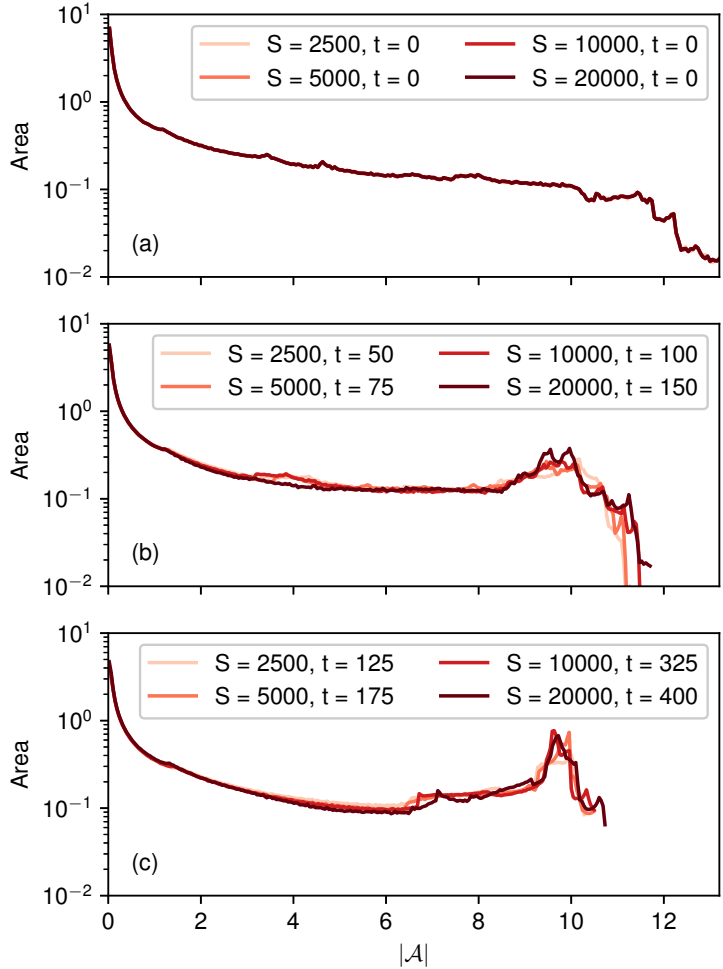}
    \caption{Histograms of $|\mathcal{A}|$ on a $1024\times 1024$ grid covering the region $\{-4<x<4, -4<y<4\}$, at times (a) $t=0$, (b) $t=t_{220}$, and (c) $t=t_{\rm max}$. The times are chosen as in Figs.~\ref{fig:timescale} and \ref{fig:flh2d}, according to the evolution of $\overline{H}$. Note that area (vertical axis) is equivalent to magnetic flux, since $B_z\equiv 1$ on $z=\pm 24$.}
    \label{fig:histo}
\end{figure}

After the relaxation, the distributions of $\mathcal{A}$ within each of the positive and negative regions are remarkably uniform, as we saw in Figure~\ref{fig:eveqn}(a/b). As a further illustration, Figure~\ref{fig:histo} shows histograms of the unsigned $|\mathcal{A}|$ distributions seen in Figure~\ref{fig:flh2d}. Each run shows the clear formation of a localized peak in the relaxed-state histogram at $|\mathcal{A}|\approx 9.5$ (Figure~\ref{fig:histo}c). After this time, which is approximately the end of the dynamical phase, the peak value is gradually reduced (like $\overline{H}$) by ohmic diffusion, at a rate dependent on $S$ but slower than the dynamical relaxation.
The topological constraint in Section~\ref{sec:top} does not explain this uniform distribution of $\mathcal{A}$ within the final flux tubes. Rather, it must arise from the physical dynamics.

\subsection{Taylor theory}

\citet{Taylor1974, Taylor1986} invoked Woltjer's earlier argument \citep{Woltjer1958} that the minimum-energy state under the constraint of conserved magnetic helicity would be a linear force-free field where ${\bf j}=\lambda_0{\bf B}$ for some constant $\lambda_0$. In our case, since there is no net helicity, the global Taylor state would be the uniform field ${\bf B}={\bf e}_z$. We have already seen how conservation of the net Poincar\'e index of the vector field $\nabla\mathcal{A}$ prevents this uniform field from being reached during the dynamical relaxation. However, we can still ask whether Taylor relaxation is operating separately within the positive and negative helicity regions. 

\begin{figure}
    \centering
    \includegraphics[width=\textwidth]{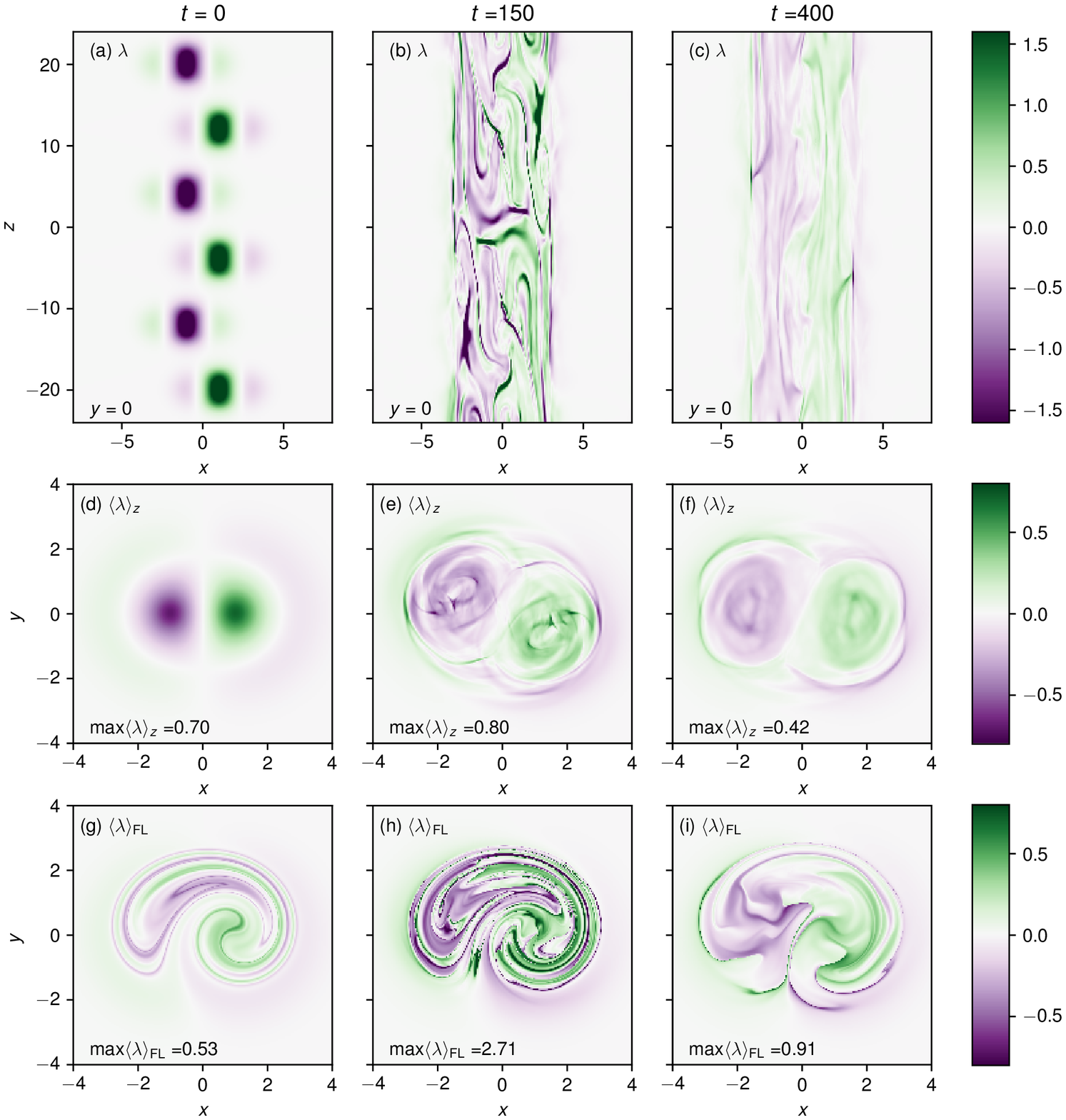}
    \caption{The force-free parameter, $\lambda={\bf j}\cdot{\bf B}/B^2$, in the $S=20\,000$ run. Panels (a-c) show slices at $y=0$, while (d-f) show means over $z$ and (g)-(i) show means along magnetic field lines (on the $z=-24$ boundary).}
    \label{fig:lambda}
\end{figure}

\begin{figure}
    \centering
    \includegraphics[width=\textwidth]{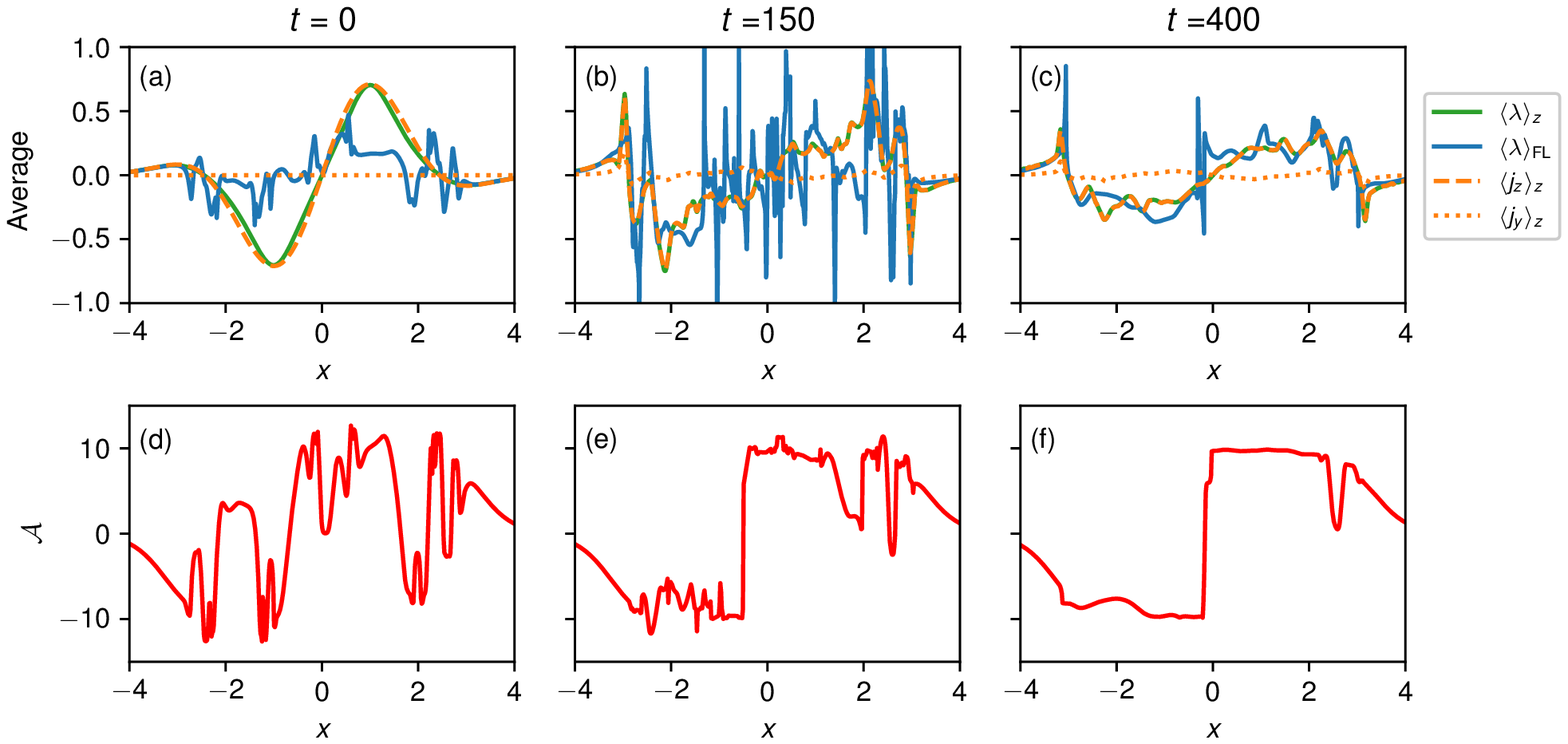}
    \caption{Comparison of the force-free parameter, $\lambda$, and field line helicity, $\mathcal{A}$ in the $S=20\,000$ run, at the same times as Figure~\ref{fig:lambda}. Panels (a)-(c) show means of $\lambda$ both in $z$ ($\langle\lambda\rangle_z$, for $y=0$) and along magnetic field lines ($\langle\lambda\rangle_{\rm FL}$, seeded at $y=0$ on $z=-24$). Also shown are means in $z$ of $j_z$ and $j_y$. For comparison, panels (d-f) show $\mathcal{A}$ at $y=0$ (seeded on the $z=-24$ boundary).}
    \label{fig:lambda1d}
\end{figure}


If we compute the profile of $\lambda = {\bf j}\cdot{\bf B}/|{\bf B}|^2$ within each flux tube, we do find some tendency for flattening. This is shown in Figures \ref{fig:lambda} and \ref{fig:lambda1d}. The flatter $\lambda$ profile is most evident at the end of the dynamical phase (Figure~\ref{fig:lambda1d}c) as compared to the initial profile (Figure~\ref{fig:lambda1d}a), although there are still quite significant variations. These variations are even more significant when $\lambda$ is averaged along magnetic field lines rather than averaged in the $z$ direction, as seen by comparing Figures \ref{fig:lambda}(f) and \ref{fig:lambda}(i), or equivalently by comparing the $\langle\lambda\rangle_z$ and $\langle\lambda\rangle_{\rm FL}$ curves in Figure \ref{fig:lambda1d}(c). Whichever averaging is used for $\lambda$, it is striking that the relaxed-state $\lambda$ profile is much less uniform than that of $\mathcal{A}$. This is evident by comparing Figure~\ref{fig:lambda1d}(c) with Figure~\ref{fig:lambda1d}(f), which shows the relaxed-state $\mathcal{A}$ in the same $y=0$ cut. Next, we propose a possible explanation.

\subsection{Relation between uniformization and Taylor theory}

Although $\lambda$ does not become completely uniform in our simulations, its tendency toward uniformity nevertheless hints at a possible explanation for the uniformization of $\mathcal{A}$. Firstly, since $B_z\approx 1$ in our field, and $|B_z| \gg \sqrt{B_x^2 + B_y^2}$, we have $\lambda\approx j_z$. Figures \ref{fig:lambda1d}(a-c) show that this holds to good approximation in our simulations.
Indeed, applying Woltjer's variational argument (minimization of magnetic energy subject to fixed magnetic helicity) to magnetic fields of the restricted form ${\bf B}=\nabla\times\big(A(r,\phi,t){\bf e}_z\big) + {\bf e}_z$ yields $j_z=\textrm{constant}$.
A flux tube that is  invariant in $z$ and has uniform $j_z$  must indeed have uniform $\mathcal{A}$.
To see this, write ${\bf B} = \nabla\times\big(A(r,\phi,t){\bf e}_z\big) + {\bf e}_z$, so that $j_z = -\nabla^2 A$. The vector potential $A$ must therefore satisfy the Poisson problem
\begin{equation}
    \nabla^2 A = -\lambda_0, \qquad A(R,\phi) = 0,
\end{equation}
where $\lambda_0$ is the uniform value of $j_z$ and $R$ is the radius of the flux tube. The unique solution must be the (regular) axisymmetric one, $A = \lambda_0(R^2 - r^2)/4$, which gives the uniform twist magnetic field ${\bf B} = (\lambda_0 r/2){\bf e}_\phi + {\bf e}_z.$ 
This indeed has uniform line helicity,
\begin{equation}
    \frac{\mathcal{A}(r,\phi)}{L_z} = \frac{{\bf A}\cdot{\bf B}}{B_z} = \frac{r}{2}B_\phi(r) + A(r) = \frac{\lambda_0R^2}{4}.
    \label{eqn:acon}
\end{equation}
For the values $\lambda_0\approx 0.25$ and $\mathcal{A}\approx 9.5$ obtained in our experiment with $L_z=48$, Equation~\eqref{eqn:acon} would predict $R\approx 1.8$, which is only a little over the actual radius of each flux tube. 

\begin{figure}
    \centering
    \includegraphics[width=\textwidth]{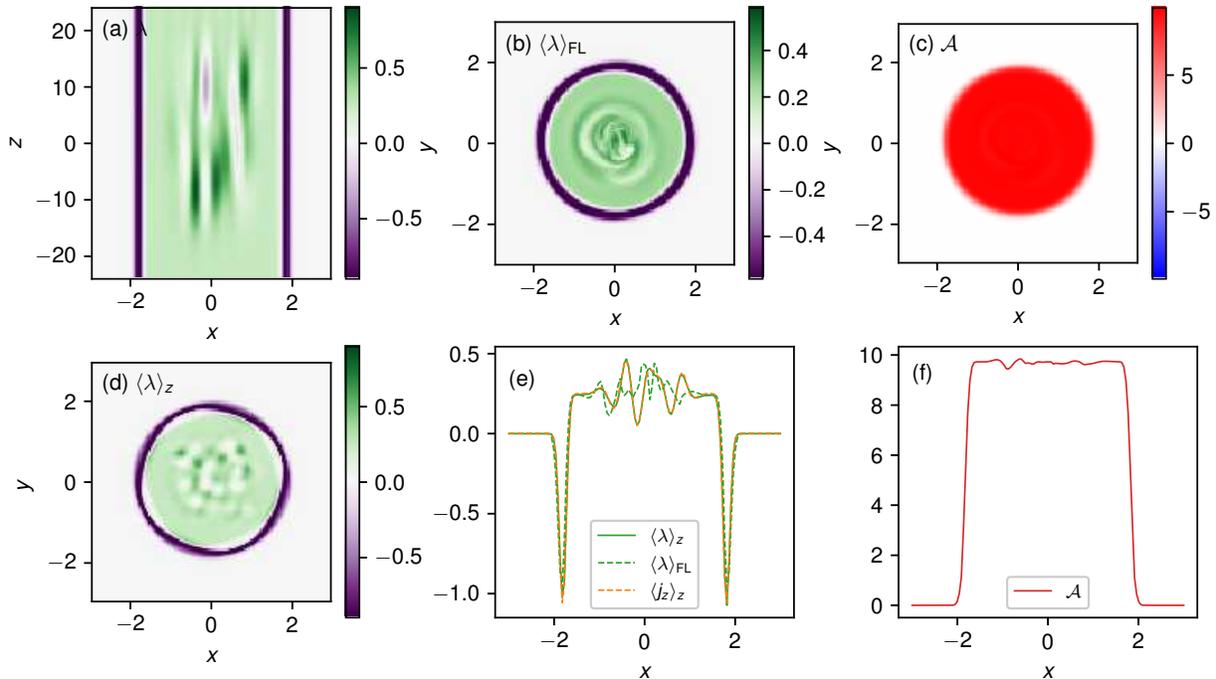}
    \caption{The simple magnetic field model to illustrate how $\mathcal{A}$ is inherently more uniform than $\lambda$ in the relaxed state. Panel (a) shows $\lambda = {\bf j}\cdot{\bf B}/B^2$ in the plane $y=0$, while panel (b) shows field line averages of $\lambda$ (in the $z=-24$ plane). Panel (c) shows line helicity also in the $z=-24$ plane, while panel (d) shows a $z$-average of $\lambda$. One-dimensional cuts at $y=0$ are shown in (e) and (f).}
    \label{fig:toy}
\end{figure}

In the simulations, it is notable that $\mathcal{A}$ shows a markedly stronger uniformity than $\lambda$, as we saw in Figure 9. This difference is natural because $\lambda$ is effectively a second derivative of $A$, whereas $\mathcal{A}$ is an integral of $A$. An alternative perspective is to observe that $\lambda$ is a locally defined quantity, while $\mathcal{A}$ can be interpreted as a non-local average over many nearby field lines \citep[cf.][]{YeatesHornig2016}. To illustrate the difference, we consider a simple analytical magnetic field model in Figure~\ref{fig:toy}, which shows $\lambda$ and $\mathcal{A}$ for the magnetic field
\begin{equation}
    {\bf B} = {\bf B}_0 + {\bf B}'.
\end{equation}
This combines an axisymmetric, uniformly-twisted background
\begin{equation}
    {\bf B}_0 = \frac{\lambda_0r}{2}\mathrm{erf}\,\left(\frac{r-R}{0.1}\right){\bf e}_\phi + {\bf e}_z,
\end{equation}
with fluctuations having the form of 100 local ``twists'',
\begin{equation}
    {\bf B}' = \sum_{i=1}^{100}b_i\exp\left(-\frac{(x-x_i)^2 + (y-y_i)^2}{(0.2)^2} - \frac{(z-z_i)^2}{6^2}\right)\left[-(y-y_i){\bf e}_x + (x-x_i){\bf e}_y\right],
    \label{eqn:bfluc}
\end{equation}
with randomly chosen strengths $b_i\in[-1,1]$ and locations $x_i,y_i\in[-R/2,R/2]$, $z_i\in[-12,12]$. This field is chosen to mimic the relaxed state in the numerical simulations, which has $j_z$ approximately constant within a tube, but with significant fluctuations about the mean. Consistent with the simulations, we set $\lambda_0=0.25$ and $R=1.8$. The distribution of $\mathcal{A}$ in Figure~\ref{fig:toy}(c/f) has been computed numerically using the same method as for the simulations.

We observe that, within our model flux tube, $\lambda$ shows significantly more fluctuations than $\mathcal{A}$, even after $\lambda$ has been averaged either in $z$ (Figure~\ref{fig:toy}d) or along magnetic field lines (Figure~\ref{fig:toy}b). The non-local nature of $\mathcal{A}$ means that the fluctuations ${\bf B}'$ tend to make equal positive and negative contributions to $\mathcal{A}$. But because $\lambda$ does not have this inherent averaging, it does not see this cancellation and maintains a significant signature from the fluctuations.

\section{Conclusion} \label{sec:con}

The numerical MHD simulations and analytical model in this paper have shown that the field line helicity $\mathcal{A}$ can add to understanding of the processes of dynamical relaxation and self organization in highly-conducting plasmas. Using direct numerical simulations, we have confirmed our earlier theoretical prediction that $\mathcal{A}$ is efficiently redistributed between field lines rather than destroyed \citep{Russell2015}. Even though our simulations are limited to relatively modest Lundquist number ranges, this phenomenon is clearly observed for all Lundquist numbers tested so far. Thus we suggest that the classic theory of Taylor relaxation could be refined by adding this ``quasi-conservation'' of field line helicity -- \textit{i.e.}, allowing it to be redistributed/exchanged between field lines but not destroyed. This remains consistent with Taylor's assumed conservation of the global helicity (because the global helicity is the weighted sum of all the individual line helicities) but it can impose additional constraints on the relaxation and therefore alter the relaxed state.

We have shown how one such constraint -- the net Poincar\'e index of the $\nabla\mathcal{A}$ pattern -- can explain why our final state comprises two oppositely-twisted magnetic flux tubes, rather than a global linear force-free magnetic field. Chen \textit{et al.} \citep{Chen2021} have recently shown that this overall topology can be predicted by a ``variational'' model that looks for the simplest possible rearrangement of the initial $\mathcal{A}$ distribution on the plane, neglecting the true turbulent dynamics altogether.

When we examine the relaxed state produced by 3D resistive MHD in finer detail, we do not find the same sub-structure within the two flux tubes that is predicted by the pure $\mathcal{A}$-rearrangement model considered by Chen \textit{et al.}\citep{Chen2021}. Our simulations with increasing Lundquist number suggest that this is not purely due to resistive decay. Rather, there is an overall 20\% increase in absolute helicity ($\overline{H}$) that appears to be independent of Lundquist number. This arises from the process of disentanglement, whereby magnetic field lines that start with portions of oppositely-signed integrand, ${\bf A}\cdot{\bf B}$, tend to reconnect and form field lines with only a single sign of ${\bf A}\cdot{\bf B}$ along their length. A similar increase in unsigned helicity has been observed in vortex reconnection \citep{Candelaresi2021}. 
Intriguingly, we found this increase to be part of a ``uniformization'' of $\mathcal{A}$ within each of the positive and negative regions. We have suggested that this arises from a Taylor-relaxation like tendency toward constant $\lambda$. The fact that line helicity is much more uniform than $\lambda$ reflects the former's more robust nature as a non-local quantity. This robustness could be useful in studying realistic turbulent plasmas. But it remains to be seen whether this uniformization of line helicity is a general behavior found in other configurations. In particular, the generality of our conclusions are limited here because the uniform-$\lambda$ and uniform-$\mathcal{A}$ states are consistent with one another owing to the relatively modest amount of magnetic helicity in our system. One can see that the two may differ in general by considering Lundquist force-free fields with differing twist (see Appendix \ref{app:l}).

Finally, the simulations in this paper considered only line-tied boundary conditions. A forthcoming paper will apply similar analysis to configurations with periodic boundary conditions (topologically toroidal). Preliminary indications suggest that the same self-organization  into opposite-helicity tubes occurs, but the definition of line helicity needs some additional care in the periodic case because there is no physical boundary delineating the end-points of individual magnetic field lines.

\begin{acknowledgments}
This work was facilitated by Leverhulme Trust grant PRG-2017-169, with additional support from Science and Technology Facilities Council (UK) consortium grants ST/N000714, ST/N000781 and ST/S000321.
\end{acknowledgments}

\section*{Data Availability Statement}

The data that support the findings of this study are available from the corresponding author upon reasonable request.

\appendix

\section{Vector potential computation} \label{app:a}

Here we describe our method for computing a vector potential $\mathbf{A}$ whose tangential components match the reference vector potential $\mathbf{A}^{\rm ref} = (-\frac12 y, \frac12 x, 0)$ on the boundaries. This reference vector potential curls to give the reference potential field $\mathbf{B}^{\rm ref}=(0,0,1)$, whose normal component, $B^{\rm ref}_{n}$, matches that of our original field, $B_n$, on all six boundaries. Our computation of $\mathbf{A}$ proceeds as follows:
\begin{enumerate}
    \item Compute a vector potential $\mathbf{A}'$ for the difference $\mathbf{B}-\mathbf{B}^{\rm ref}$ using the formulae
    \begin{align}
        A'_x(x,y,z) &= -\int_{-8}^y\Big(B_z(x,s,z) - B^{\rm ref}_{z}(x,s,z)\Big)\,\mathrm{d}s,\\
        A'_y(x,y,z) &= 0,\\
        A'_z(x,y,z) &= \int_{-8}^y\Big(B_x(x,s,z) - B^{\rm ref}_{x}(x,s,z)\Big)\,\mathrm{d}s.
    \end{align}
    \item Change gauge to $\mathbf{A}''=\mathbf{A}' + \nabla\chi$ such that $\mathbf{n}\times\mathbf{A}''=\mathbf{0}$ on all six boundaries. In fact, the tangential components of $\mathbf{A}'$ already vanish on all boundaries except for $y=8$. It suffices to take
    \begin{equation}
        \chi(x,y,z) = - \frac{y+8}{16}\int_{-8}^x A'_x(s,8,z)\,\mathrm{d}s,
    \end{equation}
    in which case
    \begin{align}
        A_x''(x,y,z) &= A_x'(x,y,z) - \frac{y+8}{16}A_x'(x,8,z),\\
        A_y''(x,y,z) &= -\frac{1}{16}\int_{-8}^x A_x'(s,8,z)\,\mathrm{d}s,\\
        A_z''(x,y,z) &= A_z'(x,y,z) - \frac{y+8}{16}A_z'(x,8,z).
    \end{align}
    \item Finally set $\mathbf{A}=\mathbf{A}'' + \mathbf{A}^{\rm ref}$.
\end{enumerate}
Since it uses only one-dimensional line integrals, this method is computationally very efficient.

\section{Line helicity in the initial configuration} \label{app:i}

The initial magnetic field given by Equations \eqref{eqn:b01}--\eqref{eqn:b03} was initially devised by \citet{WilmotSmith2009} so as to have an analytical expression for the field line mapping, in spite of that mapping's complexity. We  take advantage of this to compute the line helicity exactly for this configuration, shown in Figure~\ref{fig:index}(a). 

Recall that the magnetic field comprises six twists in a uniform background field. For our choice of parameters, the twists essentially do not overlap in the $z$ direction, so that we may derive the overall field line mapping by composing the mappings through each individual twist \citep[cf.][]{WilmotSmith2009a}. The mapping through each twist is given by $(X_{i-1},Y_{i-1}) \to (X_i, Y_i)$ where
\begin{align}
    (X_i, Y_i) &= \Big((X_{i-1} - x_i)\cos\xi_i - Y_{i-1}\sin\xi_i + x_i, \quad (X_{i-1}-x_i)\sin\xi_i + Y_{i-1}\cos\xi_i\Big),\\
    \xi_i &= 2k_i\sqrt{2\pi}\exp\left(-\frac{(X_{i-1}-x_i)^2 + Y_{i-1}^2}{2}\right).
\end{align}
Here, we have taken the mapping from $z=-\infty$ to $z=\infty$ to simplify the expressions, with no practical effect on the results  because the twists are sufficiently spaced in $z$.

The appropriate vector potential for a single twist is
\begin{equation}
    {\bf A}(x,y,z) = -\frac{y}{2}{\bf e}_x + \frac{x-x_i}{2}{\bf e}_y + \sqrt{2}k_i\exp\left(-\frac{(x-x_i)^2 + y^2}{2} - \frac{(z-z_i)^2}{4} \right){\bf e}_z + \nabla\left(\frac{x_iy}{2}\right),
\end{equation}
so that the contribution to the line helicity from this twist is
\begin{equation}
    \mathcal{A}_i(X_{i-1},Y_{i-1}) = \int_{-\infty}^\infty \frac{{\bf A}\cdot{\bf B}}{B_z}\,\mathrm{d}z = \xi_i\left(\frac{(X_{i-1}-x_i)^2 + Y_{i-1}^2}{2} + 1\right) + \frac{x_i}{2}(Y_i - Y_{i-1}).
\end{equation}
The overall line helicity for the six-twist configuration is then
\begin{align}
    &\mathcal{A}(X_0,Y_0) = \sum_{i=1}^6\mathcal{A}_i(X_{i-1},Y_{i-1}),
\end{align}
which is readily evaluated numerically.

To compute the critical points of $\mathcal{A}$ in Figure~\ref{fig:index}, zero contours of $\partial\mathcal{A}/\partial x$ and $\partial\mathcal{A}/\partial y$ were traced in the $z=0$ plane, rather than the $z=-24$ plane shown in Figures \ref{fig:flh2d} and \ref{fig:index}. The critical points remain on the same field lines and preserve their Poincar\'e index under this change of cross section. It has the advantage of reducing the sharp gradients in $\mathcal{A}$ to facilitate accurate location of the critical points. These were identified by intersections of the zero contours of $\partial\mathcal{A}/\partial x$ and $\partial\mathcal{A}/\partial y$ (equivalently, points where regions I, II, III and IV all meet simultaneously in Figure~\ref{fig:index}).

\section{Line helicity of Lundquist fields} \label{app:l}

Our purpose here is to show that Lundquist constant-$\lambda$ fields of the form
\begin{equation}
    {\bf B} = B_0\Big(J_1(\lambda_0r){\bf e}_\phi + J_0(\lambda_0r){\bf e}_z  \Big)
\end{equation}
with small enough $\lambda_0$ have very uniform field line helicity, $\mathcal{A}\approx\mathrm{constant}$. Suppose that this field is defined inside a cylinder radius $R$. The appropriate vector potential in this case is
\begin{equation}
    {\bf A} = \frac{1}{\lambda_0}\Big({\bf B} - J_0(\lambda_0R){\bf e}_z\Big),
\end{equation}
which satisfies $A_z=0$ on $r=R$ and has tangential components on the boundary with no tangential divergence \citep[cf.][]{Yeates2018}. The line helicity per unit length in $z$ for $r<R$ is therefore
\begin{equation}
    \mathcal{A}(r) = \frac{B_0}{\lambda_0}\left(\frac{J_0^2(\lambda_0r) + J_1^2(\lambda_0r)}{J_0(\lambda_0r)} - J_0(\lambda_0R)\right).
\end{equation}
We can fix $B_0$ by requiring magnetic pressure balance with a uniform field ${\bf B}={\bf e}_z$ outside the cylinder ($r>R$), which implies $B_0 = [J_0^2(\lambda_0R) + J_1^2(\lambda_0R)]^{-1/2}$.

\begin{figure}
    \centering
    \includegraphics[width=0.5\textwidth]{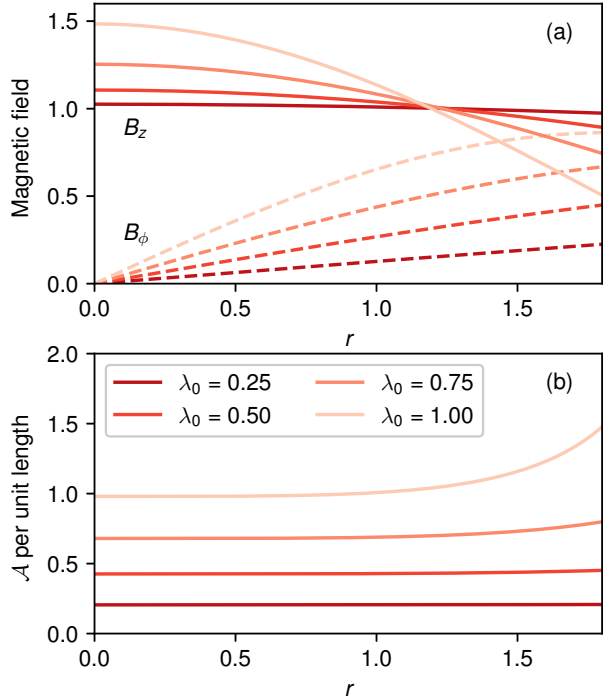}
    \caption{Radial profiles of (a) $B_\phi$ and $B_z$ and (b) $\mathcal{A}$, in four Lundquist fields with increasing $\lambda_0$ in the same cylinder with radius $R=1.8$.}
    \label{fig:lundquist}
\end{figure}

Figure \ref{fig:lundquist}(b) shows the resulting $\mathcal{A}(r)$ profiles for several values of $\lambda_0$, with corresponding $B_\phi$ and $B_z$ profiles shown in Figure \ref{fig:lundquist}(a). For small $\lambda_0$, the flux function is almost independent of $r$, including for the value $\lambda_0=0.25$ corresponding to a tube with comparable magnetic helicity to our numerical simulations. For larger $\lambda_0$, however, $\mathcal{A}$ begins to show more significant radial variation in the outer part of the tube, so that such a tube would have uniform $\lambda$ but not uniform $\mathcal{A}$. (The maximum possible $\lambda_0$ before there is a field reversal, $B_z(R)=0$, is $\approx 1.34$.)

\bibliography{relax}

\end{document}